\documentclass[12pt]{revtex4}
\usepackage{epsfig}
\begin{document}
\newcommand{\be}{\begin{equation}}
\newcommand{\ee}{\end{equation}}
\newcommand{\ba}{\begin{eqnarray}}
\newcommand{\ea}{\end{eqnarray}}

\newpage
\title{Gaussian approximation \\to the condensation of the interacting Bose
gas}
\author{Anna Okopi\'nska\\
 Institute of Physics, Pedagogical University,\\
 \'Swi\c{e}tokrzyska 15, 25-406 Kielce, Poland\\e-mail:
 okopin@fuw.edu.pl
 }

\begin{abstract}
The effective action formalism of quantum field theory is used to
study the properties of the non-relativistic interacting Bose gas.
The Gaussian approximation is formulated by calculating the
effective action to the first order of the optimized expansion. In
the homogeneous limit the method respects the Hughenholz-Pines
theorem, leading to the gapless spectrum both for excitations and
for density fluctuations. Renormalization is carried out by
adopting dimensional regularization. The results for critical
temperature are compared with that obtained in the loop expansion
and lattice calculations.
\end{abstract}
\maketitle

\section{Introduction}

A weakly interacting Bose gas was intensively studied in the
period 1947-1965 as the theoretical example of a Bose-condensed
system, described by non-relativistic QFT. Various approximation
methods have been considered~\cite{Bog}-\cite{Pethick}, achieving
a great success in describing the essential features of $^{4}He$
superfluidity. However the interactions in quantum liquids are too
strong and the weakly interacting system was in quest for a
quantitative test of many-body methods. Such a system become
accessible in 1995 after the experimental verification of the
Bose-Einstein condensation (BEC) in dilute gases in magnetic
traps~\cite{Pethick}. When adapting many-body approximations to
the trapped gases, the problem of approximation consistency
~\cite{Hoh} attracted a renewed attention~\cite{Griffin}. The
approximation scheme should preserve relations between physical
observables, arising from the symmetry of the theory (Ward
identities). For spatially uniform systems, such a relation is
implied by the Hughenholz-Pines (H-P) theorem~\cite{HP} (the
non-relativistic analogue of the Goldstone theorem). The H-P
theorem shows that the single particle spectrum of a many-body
system is gapless, if a global symmetry is spontaneously broken.
This imposes strong constraints on approximations. The simplest
Bogoliubov approximation~\cite{Bog} fulfils the constraints, but
including the first order corrections to propagators violates the
H-P theorem. Only the second order approximation, developed by
Beliaev~\cite{Bel} and extended to finite temperature by Popov, is
fully consistent. In the self-consistent Hartree-Fock-Bogoliubov
approximation the H-P is violated: there is a gap in the single
particle spectrum although the density fluctuations are
gapless~\cite{Griffin}.

The Bose condensed systems are usually studied within field
theoretical techniques involving the particle Hamiltonian. The
Lagrangian formulation is more convenient for discussing the
general issues, such as symmetries, conservation laws, and Ward
identities. Upon straightforward extension to finite temperature
the grand canonical thermodynamic potential is obtained as a
function of the superfluid order parameter. However, the results
obtained in the Lagrangian approach are also controversial:
Toyoda~\cite{Toy} claimed the one-loop effective potential at
finite temperature to be consistent with the H-P theorem, but
later an opposite statement has been made~\cite{Rav}. Methods of
self-consistent re-summation have been also
discussed~\cite{Rav,Stoof} but the consistency with the H-P
theorem is unclear. We show that the effective action provides a
useful tool to construct consistent approximations.

The outline of our work is as follows. In section~\ref{ea} the
effective action formalism is introduced using path-integrals for
generating functionals of non-relativistic theory. We show that
the H-P theorem can be formulated in a form of a simple criterium,
which is useful for checking approximations consistency. In
section~\ref{loop} we study the loop expansion for the effective
action. The one-loop result gives the Beliaev-Popov approximation
at finite temperature. In section~\ref{oe} we extend the method of
the optimized expansion~\cite{AO} to the effective action of
non-relativistic theory. The self-consistent Gaussian
approximation, obtained in the first order, fulfils the H-P
theorem. Results for the critical temperature and conclusions are
summarized in Section~\ref{con}.
\section{Effective action}\label{ea}
The system of many spinless atoms interacting by two-body forces
is described by a complex scalar field $\Phi(\textbf{r},t)$ with
the Lagrangian density\ba
\emph{L}[\Phi]=\frac{i}{2}\left(\Phi^{*}(\textbf{r},t)\frac{\partial\Phi
(\textbf{r},t) }{\partial t }
-\frac{\partial\Phi^{*}(\textbf{r},t)}{\partial t
}\Phi(\textbf{r},t)\right)-
\frac{\hbar^{2}}{2m}\nabla\Phi^{*}(\textbf{r},t)\cdot\nabla\Phi(\textbf{r},t)
\nonumber\\-V_{ext}(\textbf{r})\Phi^{*}(\textbf{r},t)
\Phi(\textbf{r},t)-\frac{1}{2}\!\int \!d^{3}r'
\left(\Phi^{*}(\textbf{r},t)
\Phi(\textbf{r},t)U(\textbf{r}\!-\!\textbf{r'})\Phi^{*}(\textbf{r'},t)
\Phi(\textbf{r'},t)\right),\label{Lcl} \ea where
$U(\textbf{r}-\textbf{r'})$ represents the two-body interatomic
potential, and $V_{ext}(\textbf{r})$ the external potential of the
trap. The Lagrangian is invariant under the global U(1)
transformation\be (\Phi,\Phi^{*})\rightarrow
(e^{i\alpha}\Phi,e^{-i\alpha}\Phi^{*}),\label{inv}\ee where
$\alpha$ is a constant phase. Since BEC takes place at very low
energies, the interatomic potential can be approximated by the
local potential $U(\textbf{r}-\textbf{r'})=
2\lambda\delta(\textbf{r}-\textbf{r'})$ with
$\lambda=\frac{2\pi\hbar^{2}a}{m}$ related to the scattering
length $a$. The Lagrangian simplifies to the form \ba
\emph{L}[\Phi]&=&\frac{i}{2}\left(\Phi^{*}(\textbf{r},t)\frac{\partial\Phi
(\textbf{r},t) }{\partial t }
-\frac{\partial\Phi^{*}(\textbf{r},t)}{\partial t
}\Phi(\textbf{r},t)\right)-
\frac{\hbar^{2}}{2m}\nabla\Phi^{*}(\textbf{r},t)\cdot\nabla\Phi(\textbf{r},t)
\nonumber\\&-&V_{ext}(\textbf{r})\Phi^{*}(\textbf{r},t)
\Phi(\textbf{r},t)-\lambda \left(\Phi^{*}(\textbf{r},t)
\Phi(\textbf{r},t)\right)^2.\label{Lcl0} \ea The grand canonical
system at the temperature $T$ in the imaginary time formalism is
defined by the Wick's rotated Lagrangian, \ba
\emph{L}_{\mu}[\Phi]=\Phi^{*}(\textbf{r},\tau)\left(-\frac{\partial}{\partial\tau}+
\frac{\hbar^{2}}{2m}\nabla^2-V_{ext}(\textbf{r})+\mu\right)\Phi
(\textbf{r},\tau) -\lambda \left(\Phi^{*}(\textbf{r},\tau)
\Phi(\textbf{r},\tau)\right)^2,\label{LEu} \ea where the chemical
potential $\mu$ is introduced in order to consider states with an
indefinite number of particles. Its value is to be adjusted so
that the expectation value of the number operator is equal to
$\textsl{N}$, corresponding to a fixed particle density,
$n=\frac{\textsl{N}}{\int d^{3}r}.$ The Euclidean generating
functional is defined as a path integral,
\begin{equation}
Z[J]=\int\! D\Phi D\Phi^{*}\,e^{-\frac{1}{\hbar}\int_{0}^{\beta}
d\tau\int d^{3}r\left [\emph{L}_{\mu}[\Phi]
+\!J^{*}(\textbf{r},\tau)\Phi(\textbf{r},\tau)+\!
J(\textbf{r},\tau)\Phi^{*}(\textbf{r},\tau)\right]}\label{Z}
\end{equation}
over the functions $\Phi(\textbf{r},\tau)$ with a period
$\beta=\frac{\hbar}{k_{B}T}$ in $\tau$. The partition function at
thermal equilibrium is given by $Z[0]$.  The generating functional
for connected Green's functions  $ W[J]=\ln Z[J]$. The effective
action functional is given by the Legendre transform
\begin{equation}
\Gamma[\Phi]=W[J]-\int dx J(x) \Phi^{*}(x)-\int dx
J^{*}(x)\Phi(x), \label{Gam}
\end{equation}
where $x\!=\!(\tau,\textbf{r})$. The background field
$\Phi(x)\!=\!\frac{\delta W}{\delta J(x)}\!=\!\langle
\widehat{\Phi}(x)\rangle_{J}$ is a vacuum expectation of the
quantum field operator in the presence of external source,
$J(x)=\frac{1}{\sqrt{2}}\left(j_{1}(x)+ij_{2}(x)\right)$.
Performing the calculation we will use the two real components of
the complex field
$\Phi(x)=\frac{1}{\sqrt{2}}\left(\phi_{1}(x)+i\phi_{2}(x)\right)$
as independent variables.  As a Legendre transform the effective
action fulfils \be \frac{\delta \Gamma}{\delta
\phi_{i}(x)}=-j_{i}(x)~~~~\mbox{where}~~~~i=1,2.\ee The physical
value of the background field $\Phi^{(0)}(x)$, corresponding to
$J(x)=0$ is thus determined by the stationarity equations
\be\left.\frac{\delta \Gamma}{\delta
\phi_{i}(x)}\right|_{\Phi^{(0)}(x)}=0~~~~~~~~~~~~\mbox{where}~~~~i=1,2.\ee
Green's functions are obtained by functional differentiation of
generating functionals. The one-particle Green's function
(propagator) matrix reads
$$G_{ij}(x,y)\!=\frac{1}{Z{[0]}}\!\left.\frac{\delta^{2} Z}{\delta
j_{i}(x)\delta
j_{j}(y)}\right|_{J\!=\!0}\!=\!\left.\frac{\delta^{2} W}{\delta
j_{i}(x)\delta j_{j}(y)}\right|_{J\!=\!0}.$$ By differentiation of
the effective action the one-particle irreducible Green's
functions (proper vertices) are generated. The proper vertex \be
\Gamma_{ij}(x,y)=\left.\frac{\delta^2\Gamma}{\delta
  \phi_{i}(x)\phi_{j}(y)}\right|_{\Phi=\Phi^{(0)}}=\left[\left.\frac{\delta^2W}{\delta
  j_{i}(x)j_{j}(y)}\right|_{J=0}\right]^{-1}=G_{ij}^{-1}(x,y).\ee One-particle excitations,
  related to the poles of the full propagator, are determined by zero modes of the
Fourier transform $\left.\Gamma_{ij}(p)=\int dx
e^{-ip(x-y)}\frac{\delta^2\Gamma}{\delta
  \phi_{i}(x)\phi_{j}(y)}\right|_{\Phi=\Phi^{(0)}},$ where $p$ stands for
$(\omega,\textbf{p})$. The matrix can be written in the form:
$$\Gamma(p)=\left(\begin{array}{cc}
  \frac{p^2}{2m}-\mu+\Pi_{11} &\omega+\Pi_{12} \\
  -\omega+\Pi_{21} & \frac{p^2}{2m}-\mu+\Pi_{22}
\end{array}\right)$$
where
$\Pi_{11}\!=\Sigma_{11}\!+\frac{1}{2}\left(\Sigma_{12}\!+\Sigma_{12}^{*}\right)$,
$\Pi_{22}\!=\Sigma_{11}\!-\frac{1}{2}\left(\Sigma_{12}\!+\Sigma_{12}^{*}\right)$
and
$\Pi_{12}\!=\frac{i}{2}\left(\Sigma_{12}\!-\Sigma_{12}^{*}\right)$
with $\Sigma_{11}\!=\!\Sigma_{22}$ being the normal self-energy
and $\Sigma_{12}\!=\!\Sigma_{21}^{*}$ the \mbox{anomalous one.}

In the following we study a homogeneous system,
$V_{ext}(\textbf{r})\rightarrow 0$, when the background field
$\Phi^{(0)}(x)=\Phi^{(0)}$. In this case the effective potential
\begin{equation}
V(\Phi)=-\frac{\left.\Gamma[\Phi]\right|_{\Phi(x)=\Phi
=const}}{\beta \int\! d^{3}r}
\end{equation}
is an useful tool, since $\Phi^{(0)}$ can be determined by the
stationarity equation \be \left.\frac{dV}{d
\phi_{1}}\right|_{\Phi^{(0)}}=\left.\frac{dV}{d
\phi_{2}}\right|_{\Phi^{(0)}}=0.\ee The H-P theorem can be easily
demonstrated in the effective action approach. Because of the
invariance of $\emph{L}_{\mu}[\phi_{1},\phi_{2}]$ under the
transformation~(\ref{inv}), whose infinitesimal version is given
by \ba\left(\begin{array}{c}
  \phi_{1}\\
  \phi_{2}
\end{array}\right)\rightarrow\left(\begin{array}{cc}
  1 & -\alpha \\
  \alpha & 1
\end{array}\right)\left(\begin{array}{c}
 \phi_{1}\\
  \phi_{2}
\end{array}\right),
\ea the generating functional $W[j_{1},j_{2}]$ is invariant
under\ba\left(\begin{array}{c}
  j_{1}\\
  j_{2}
\end{array}\right)\rightarrow\left(\begin{array}{cc}
  1 & \alpha \\
  -\alpha & 1
\end{array}\right)\left(\begin{array}{c}
 j_{1}\\
  j_{2}
\end{array}\right).\ea
This implies
\begin{equation}\label{dw}
  \int dx \left(\frac{\delta W}{\delta j_{1}} j_{2}-\frac{\delta W}{\delta j_{2}}
  j_{1}\right)=\int dx \left(- \phi_{1}\frac{\delta \Gamma}{\delta
  \phi_{2}}+
   \phi_{2}\frac{\delta \Gamma}{\delta \phi_{1}}\right)=0.
 \end{equation}
Taking derivative over $\phi_{2}(y)$ at
$\Phi^{(0)}\!=\!\frac{1}{\sqrt{2}}(\phi_{1}^{(0)},0)$ which
fulfils the symmetry breaking condition $
   \left.\frac{\delta \Gamma}{\delta
   \phi_{1}(x)}\right|_{\Phi^{(0)}}=0,
$  we obtain \ba\label{dww}
 && \int dx \left(-\phi_{1}(x)\left.\frac{\delta^2\Gamma}{\delta
  \phi_{2}(x)\phi_{2}(y)}\right|_{\Phi^{(0)}}+
   \left.\frac{\delta \Gamma}{\delta \phi_{1}(x)}\right|_{\Phi^{(0)}}+
   \left.\phi_{2}(x)\frac{\delta ^2\Gamma}{\delta \phi_{1}(x)\delta \phi_{2}(y)}
   \right|_{\Phi^{(0)}}\right)\nonumber\\&&=-\phi_{1}^{(0)}\int dx
   \left.\frac{\delta^2\Gamma}{\delta
  \phi_{2}(x)\phi_{2}(y)}\right|_{\Phi^{(0)}}=
  -\phi_{1}^{(0)}\Gamma^{22}(p=0)=0.
 \ea
Since $\phi_{1}^{(0)}\neq0$, this means that
$\Gamma^{22}(\omega=0,\overrightarrow{k}=0)=0$, there is therefore
a zero-frequency excitation. The H-P theorem can be expressed in
the form \be
\Gamma^{22}(\omega=0,\overrightarrow{k}=0)\!=\!\left.\frac{d^{2}V}{d\phi_{2}^2}\right|_{\Phi^{(0)}}=0.\ee
It is easy to observe that the gapless spectrum is just a
consequence of the fact that the effective potential is a function
of $|\Phi|^2$ and doesn't depend on $\phi_{1}$ and $\phi_{2}$
separately. This provides a useful criterium for the consistency
of an approximation with the H-P theorem.

For interacting atoms $(a\!\neq\!0)$ the effective cannot be
calculated exactly, so one resorts to approximations. The
advantage of formulating the approximation for the effective
action lies in the fact that all the Green's functions are
obtained in a consistent way, through functional differentiation.
\section{The loop expansion} \label{loop}
The loop expansion is generated by calculating the path integral
for $Z[J]$ by the steepest descent method. Upon Legendre
transformation the effective action is obtained as a series in
$\hbar$, whose power indicates the number of loops. The 1-loop
effective action
\ba\Gamma^{1-loop}[\Phi]\!\!\!&\!=\!&\!\!\!\!\int\!dx\!
\left[\!\frac{i}{2}\phi_{1}\frac{\partial \phi_{2}}{\partial
\tau}\!-\frac{i}{2}\phi_{2}\frac{\partial \phi_{1}}{\partial \tau}
\!-\!
\frac{1}{2}\phi_{1}(\frac{\!\nabla^2\!\!}{2m}+\!\mu)\phi_{1}\!\!-\!
\frac{1}{2}\phi_{2}(\frac{\!\nabla^2\!\!}{2m}+\!\mu)\phi_{2}\right.
\nonumber\\&+&\left.\!\frac{\lambda}{4}\left(\phi_{1}^{2}\!+\!
\phi_{2}^{2}\right)^2\!\right]+\frac{\hbar}{2}TrLn
M[\Phi],\label{LEea} \ea where $
M[\Phi]\!=\!\left[\!\begin{array}{cc}
   -\frac{\nabla^2}{2m} -\mu+3\lambda\phi_{1}^2(x)+
   \lambda\phi_{2}^2(x)&  i\frac{\partial}{\partial\tau}+2\lambda \phi_{1}(x)\phi_{2}(x)\\
   -i\frac{\partial}{\partial\tau}+2\lambda
   \phi_{1}(x)\phi_{2}(x)
    & -\frac{\nabla^2}{2m}-\mu+\lambda \phi_{1}^2(x)+3\lambda
    \phi_{2}^2(x)
\end{array}\!\right],$\\\\gives the Beliaev-Popov
approximation at finite temperature. Setting
\mbox{$\Phi(x)\!=\!\Phi$} yields the Toyoda's~\cite{Toy} result
for 1-loop finite temperature effective potential\be
V^{1-loop}(|\Phi|^2)\!=-\mu|\Phi|^{2}+\!\lambda |\Phi|^{4}+\hbar
I_{1}(\mu,|\Phi|^2)\label{LEep}\ee  where \be
I_{1}(\mu,|\Phi|^2)\!=\!\int\!\frac{d^3k}{(2\pi)^3}\!\left[
\frac{1}{2} \omega_{k}\!+\frac{1}{\beta} ln\left(1-e^{-\beta
\omega_{k}}\!\right)\right]\label{I1}\ee and
 $\omega_{k}\!=\!\sqrt{\left(\frac{k^2}{2m}-\mu+4\lambda|\Phi|^2\right)^2-4\lambda^2|\Phi|^4}.$
The H-P theorem is respected in the 1-loop approximation, since
the effective potential depends only on $|\phi|^2$. The false
statement in Ref.~\cite{Rav} was due to wrong interpretation of
excitation energy. One can verify explicitly that
$\left.\Gamma\!_{22}(\!p\!=\!0\!)\!=\!\frac{d^2
V^{1-loop}}{d\phi_{2}^2}\right|_{\Phi^{(0)}}\!\!=\!0$, if
$\Phi^{(0)}\!=\!\frac{1}{\sqrt{2}}(\phi_{1}^{(0)},0)$ is a
 solution
 to the
stationarity equation \be
\left.\frac{dV^{1-loop}}{d\phi_{1}}\right|_{\Phi^{(0)}}=\left[-\mu+2\lambda|\Phi^{(0)}|^2+\hbar
\lambda J_{0}(\mu,|\Phi^{(0)}|^2)
\right]\phi_{1}^{(0)}\!\!=\!0,\ee where \be
J_{0}(\mu,|\Phi|^2)\!=\!\!\int\!\!\frac{d^3k}{(2\pi)^3}
\frac{\frac{k^2}{2m} - \mu +
3\lambda|\Phi|^2}{\omega_{k}}\left(1\!+\!2
n_{B}(\omega_{k})\!\right)\label{J0}\ee and the Bose-Einstein
distribution function
$n_{B}(\omega)\!=\!\frac{1}{(e^{\beta\omega}-1)}$.

The chemical potential $\mu$ can be eliminated in favor of the
particle number density, using the relation\be
n=-\left.\frac{dV}{d\mu}\right|_{\Phi^{(0)}}=|\Phi_{0}|^2+ \hbar
I_{0}(\mu,|\Phi_{0}|^2),\ee where \be
I_{0}(\mu,|\Phi|^2)\!=\!\int\frac{d^3k}{(2\pi)^3}
\frac{\frac{k^2}{2m}-\mu+4\lambda|\Phi|^2}{\omega_{k}}\left(1\!+\!2
n_{B}(\omega_{k})\right).\label{I0}\ee  Renormalization can be
carried out by dimensional regularization. Denoting the particle
density in condensed state $|\Phi_{0}|^2$ by $n_{0}$ and observing
that in the leading order $n_{0}\!=\!n$, the particle density to
the \mbox{order $\hbar$} can be written
 as
\be n=n_{0}+ \int\frac{d^3k}{(2\pi)^3}
\frac{\frac{k^2}{2m}+2\lambda n }{\omega_{k}^{Bog}}\left(1\!+\!2
n_{B}(\omega_{k}^{Bog})\right),\label{den}\ee where Bogoliubov
frequency
$\omega_{k}^{Bog}=\sqrt{\frac{k^2}{2m}(\frac{k^2}{2m}+4\lambda
n)}$. This expression has been used~\cite{Benson} to calculate the
condensate depletion\be n_{0}=n-\frac{8(a n)^{3/2}}{3\sqrt{\pi}}
-\left(\frac{m}{2\pi\beta}\right)^{2/3}\!\!
\left[\zeta(\!3/2\!)-2\pi \sqrt{\frac{2\beta a
n}{m}}+O(\beta)\right].\ee The depletion and other physical
quantities (energy density, pressure, $\ldots$) derived from
1-loop effective potential
 are in agreement
with the Lee and Yang results~\cite{LY} to the lowest order in $a
n^{1/3}$. However, the 1-loop results are dubious in the vicinity
of the phase transition, since the higher-loop contributions are
significant at such a temperature.
\section{The optimized expansion} \label{oe}
The optimized expansion~\cite{AO} consists in introducing an
arbitrary parameter $\Omega$ into the Lagrangian density \ba
&&\emph{L}_{\mu}^{\epsilon}[\Phi,\Omega]=\Phi^{*}(\textbf{r},\tau)\left(-\frac{\partial}{\partial\tau}+
\frac{\hbar^{2}}{2m}\nabla^2+\Omega\right)\Phi
(\textbf{r},\tau)\nonumber\\
&&+\epsilon\left[\Phi^{*}(\textbf{r},\tau)\left(\mu-\Omega\right)\Phi
(\textbf{r},\tau) -\lambda \left(\Phi^{*}(\textbf{r},\tau)
\Phi(\textbf{r},\tau)\right)^2\right].\label{OEu} \ea For
$\epsilon\!=\!1$ the dependence on $\Omega$ cancels and the
modified Lagrangian coincides with the original one~(\ref{LEu}).
Calculating Z[J] and performing the Legendre transform yields the
effective action as a series in a formal parameter $\epsilon$. The
exact result would not depend on $\Omega$, however such a
dependence appears in the $n$-th order truncation,
$\Gamma^{(n)}[\Phi,\Omega]$, obtained after setting $\epsilon=1$.
We exploit this freedom by choosing the value of $\Omega$, which
fulfils the minimal sensitivity requirement
\begin{equation}
\left.\frac{\delta\Gamma^{(n)}}{\delta \Omega
}\right|_{\Omega^{opt}}= 0,\label{sta}
\end{equation}
in the given order approximation
$\Gamma^{(n)}[\Phi,\Omega^{opt}]$. The optimal value of $\Omega$
changes from order to order, improving the convergence properties
of the scheme. The approach is equivalent to a systematic
re-summation of the perturbation series.

Gaussian approximation to the effective action is obtained in the
first order of the optimized expansion. This yields the Gaussian
effective potential \ba
V^{(1)}[\Phi,\Omega]\!=-\mu|\Phi|^{2}+\!\lambda |\Phi|^{4}+
I_{1}(\Omega,|\Phi|^2)+(\mu-\Omega)I_{0}(\Omega,|\Phi|^2)+3\lambda
I_{0}^2(\Omega,|\Phi|^2)\ea with self-consistency condition\be
\frac{dV^{(1)}}{d\Omega}\!=\![(\mu-\Omega)+6\lambda
I_{0}(\Omega,|\Phi|^2)]\frac{dI_{0}}{d\Omega}=0,\label{sel}\ee
where $I_{1}$ and $I_{0}$ are defined by~(\ref{I1}) and
(\ref{I0}), respectively. Since both the effective potential and
the optimization condition depend only on $|\Phi|^2$, one clearly
sees that the H-P theorem is respected. The symmetry is
spontaneously broken at $|\Phi|^2=n_{0}$ which renders the
effective potential stationary:\be
\left.\frac{dV^{(1)}}{d|\Phi|^2}\right|_{|\Phi|^2\!=\!n_{0}}
\!\!\!\!\!\!=\!-\mu+\!2\lambda n_{0}^{2}+\lambda
J_{0}(\Omega,n_{0})+[(\mu-\Omega)+6\lambda
I_{0}(\Omega,n_{0})]\left.\frac{dI_{0}(\Omega,|\Phi|^2)}{d|\Phi|^2}
\right|_{|\Phi|^2\!=\!n_{0}}\!\!\!\!\!\!=\!0\label{gap}\ee with
$J_{0}$ given by~(\ref{J0}). The particle density in our
approximation reads \be
n=\!-\frac{dV^{(1)}}{d\mu}\!=\!n_{0}+I_{0}(\Omega,n_{0})\label{n}\ee
Upon eliminating $\mu$ by (\ref{sel}), the Eqs.\ref{gap} and
\ref{n} simplify to
\ba n&=&\!n_{0}+ I_{0}(\Omega,n_{0})\nonumber\\
\Omega&=&\!2\lambda n_{0}^{2}+\lambda J_{0}(\Omega,n_{0})+6\lambda
I_{0}(\Omega,n_{0})\label{self}\ea which, after numerically
eliminating $\Omega$, determine $n_{0}(n,\beta)$.
\section{Results and conclusions} \label{con}
The critical temperature for BEC, $T_{c}$, is the temperature
below which the symmetry is spontaneously broken. Its value can be
calculated from the condition that $|\Phi|^2\!=\!n_{0}\!=\!0$ at
the phase transition. For the ideal gas $
T_{c}^{id}=\left(\frac{2\pi}{m}\right)\left(\frac{n}{\zeta(\frac{3}{2})}
\right)^\frac{2}{3}$, but calculating the shift of the critical
temperature in the presence of interactions \be\Delta
T_{c}=\frac{T_{c}-T_{c}^{id}}{T_{c}^{id}}\ee generates
controversy, even for the dilute gas. Different powers of the
leading behavior in $an^{1/3}$ with different coefficients have
been reported by various authors~\cite{Arnold}. $T_{c}$ is usually
derived from the effective 3-dimensional theory, arguing that only
the zero Matsubara modes determine the critical behavior. In this
approximation the leading behavior $\Delta T_{c}\approx 1.32
an^{1/3}$~\cite{lattice} and next-to-leading order corrections
have been determined~\cite{Arnold}.

Here we show the approximate results for $T_{c}$ derived in the
original theory in $(3+1)$-dimensions. The 1-loop approximation
for $T_{c}$ has been calculated~\cite{Kleinert} by setting the
value of the background field $n_{0}$ to zero in Eq.~\ref{den},
which yields the equation \be n= I_{0}(2\lambda n,n),\ee to be
solved numerically. We calculate the Gaussian approximation for
$T_{c}$ by setting $n_{0}=0$ in
Eq.~\ref{self}, which results in the coupled pair of equations\ba n&=&I_{0}(\Omega,0)\nonumber\\
\Omega&=&\lambda J_{0}(\Omega,0)+6\lambda
I_{0}(\Omega,0).\label{self}\ea

In Fig.1 the numerical results for $T_{c}$ are compared with the
results derived in the dimensionally reduced
theory~\cite{lattice,Arnold}. For the experimental evaluation of
$T_{c}$ the accuracy achieved in homogeneous systems is still
insufficient. The theoretical results differ much, even at small
values of $an^{1/3}$. It is interesting to observe that the
leading behavior in 1-loop approximation is of the form $\Delta
T_{c}\approx\frac{4\sqrt{\pi}}{3}
\frac{\sqrt{an^{1/3}}}{[\zeta(3/2)]^{2/3}}$ (with the sign
opposite to that obtained by Toyoda~\cite{Toy}), while the
Gaussian approximation gives $\Delta
T_{c}\approx\frac{4\sqrt{2\pi}}{3}
\frac{\sqrt{an^{1/3}}}{[\zeta(3/2)]^{2/3}}$ with the coefficient
$\sqrt{2}$ times larger than 1-loop result. In both cases the
square root dependence on $an^{1/3}$ is obtained, similar to Lee
and Yang result~\cite{LY}, but in difference with the linear
dependence derived from the dimensionally reduced theory. The
behavior of $T_{c}$ in the Gaussian approximation at larger values
of $an^{1/3}$ becomes qualitatively different from the 1-loop
behavior and there is no sign of the reentrant phase transition
suggested~\cite{Kleinert} by the 1-loop approximation. It would be
interesting to investigate this issue in higher order
approximations. Both the loop expansion, as well as the optimized
one, offer a systematic and consistent way for such a study,
without violating the H-P theorem.

\begin{figure}[h]
\setlength{\unitlength}{1cm}
\centerline{\epsfxsize=7cm\epsfbox{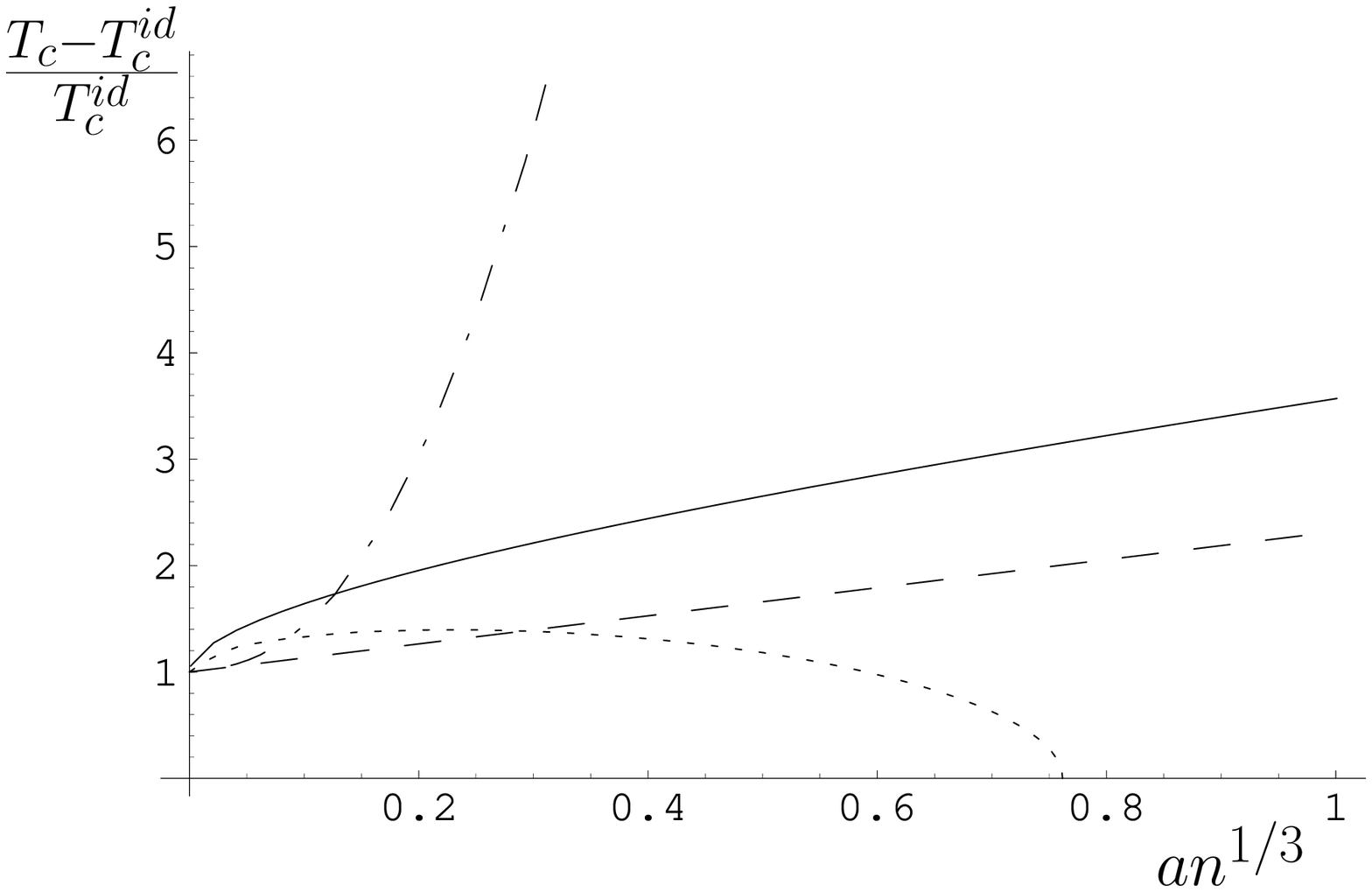}} \caption{The critical
temperature for BEC in the Gaussian
approxi\-ma\-tion(\textit{solid line}), compared with the 1-loop
result~\cite{Kleinert} (\textit{dotted line}), and the results of
the 3-dimensional theory:  the leading order in
$an^{1/3}$~\cite{lattice} (\textit {dashed line}), and next-to
leading order~\cite{Arnold} (\textit {dotted-dashed line}).}
\end{figure}
\section{Acknowledgment}The work was partially supported by the
Committee for Scientific Research under Grant No.2P03B7522.

\end{document}